\documentclass[twocolumn,showpacs,preprintnumbers,amsmath,amssymb]{revtex4}
\usepackage{graphicx}
\usepackage{dcolumn}
\usepackage{bm}
\usepackage[tight]{subfigure}
\usepackage{amsmath}
\usepackage{verbatim}
\usepackage{color}

\begin{document}

\title{Scheme to measure Majorana fermion lifetimes using a quantum dot}

\author{Martin Leijnse}
\author{Karsten Flensberg}

\affiliation{
  Nano-Science Center \& Niels Bohr Institute,
  University of Copenhagen,
  2100~Copenhagen \O, Denmark 
}

\begin{abstract}
We propose a setup to measure the lifetime of the parity of a pair of Majorana bound
states. The proposed experiment has one edge Majorana state tunnel coupled to a
quantum dot, which in turn is coupled to a metallic electrode. When the Majorana
Fermions overlap, even a small relaxation rate qualitatively changes the non-linear
transport spectrum, and for strong overlap the lifetime can be read off directly
from the height of a current peak. This is important for the usage of Majorana
Fermions as a platform for topological quantum computing, where the parity
relaxation is a limiting factor.

\end{abstract}
\maketitle
Topological superconductors are currently attracting massive interest, partly due to the Majorana bound states 
(MBS) which form at edges or vortices of such systems. 
Majorana Fermions are non-Abelian anyons~\cite{Stern10rev}, 
meaning that particle exchanges are non-trivial operations which in general do not commute and 
can be used to perform quantum computational operations, called topological quantum computing~\cite{Nayak08rev}. 
Non-Abelian statistics has been predicted for the quasi-particle excitations of the 
$\nu = 5/2$ fractional quantum Hall state~\cite{Moore91}, 
and appear as a consequence of $p$-wave type pairing~\cite{Kraus09}, 
and such superconductors should host MBS in vortices. 
Recently, it was realized that $p$-wave like pairing may also occur in topological 
insulators~\cite{Fu08}, and even in ordinary semiconductors with strong spin-orbit coupling~\cite{Sau10, Alicea10, Akhmerov11}, 
when brought into proximity with an $s$-wave superconductor. 

A particularly simple system is a one-dimensional semiconducting 
wire with strong spin-orbit coupling, brought into proximity with an $s$-wave superconductor~\cite{Oreg10, Lutchyn10}. 
The original idea of MBS in wires is much older~\cite{Kitaev01}, 
the more recent proposals being possible experimental realizations of that model system.
Particle exchanges (braiding)
could be accomplished by crossing two or more wires~\cite{Sau10b, Alicea10b}.
However, the first step would be to verify the existence of MBS,
which could e.g., be achieved by tunnel spectroscopy~\cite{Bolech07, Law09}. The presence of a MBS gives rise to a 
characteristic zero-bias conductance peak,
while a more complicated peak-structure arises if many MBS are coupled to each other~\cite{Flensberg10}. 

The main advantage of topological quantum computing
is the robustness against decoherence: the computational basis consists of pairs of 
MBS which are spatially separated and the state normally cannot be modified by perturbations which do not 
couple simultaneously to more than one MBS~\cite{Nayak08rev}. 
However, perturbations which change the parity degree of freedom of the superconductor \emph{can} change 
the state of the Majorana system and thus lead to decoherence~\cite{Nayak08rev}. The presence of such parity-changing processes, 
called quasiparticle poisoning~\cite{Mannik04, Aumentado04}, is in fact a well-known problem in 
superconducting charge qubits.
In order to determine whether Majorana bound states in different topological materials are actually suitable 
for topological quantum computing, it is therefore crucial to be able to measure the lifetime of the parity degree
of freedom. Here we show how this can be done in an experimentally rather simple transport setup, 
which does not require braiding or interferometry.

\begin{figure}[t!]
  \includegraphics[height=0.3\linewidth]{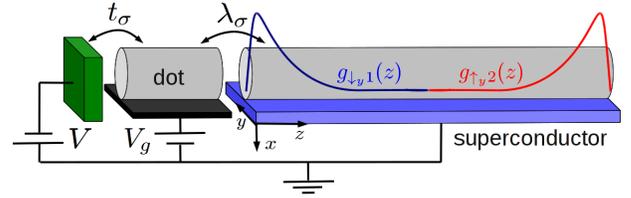}	
	\caption{\label{fig:1}
	(Color online) Sketch of the proposed setup. 
	An $s$-wave superconductor induces superconductivity in a semiconducting wire with strong spin-orbit coupling.
        A magnetic field can then induce a topological phase where MBS are formed at the ends 
	of the wire. A quantum dot is formed e.g., by separating a short segment from the main wire by a tunnel barrier,
        and it is tunnel coupled to a metallic normal lead. 
	}
\end{figure}

The parity lifetime cannot be measured in a normal metal--MBS tunnel junction as studied in 
Refs.~\cite{Bolech07, Law09, Flensberg10}, and we consider instead a setup including a quantum dot between 
the normal metal and the topological superconductor, see Fig.~\ref{fig:1}. 
We focus here on a wire-type topological superconductor, which seems experimentally most 
attractive, but the main conclusions are independent of such details.
Systems of several coupled quantum dots and MBS have been suggested as probes of the non-locality of 
Majorana Fermions~\cite{Tewari08} and very recently as an alternative way to perform non-Abelian rotations 
within the degenerate Majorana groundstate manifold~\cite{Flensberg10b}.
However, the non-equilibrium transport properties 
of a normal metal--quantum dot--MBS junction have to our knowledge not been 
investigated previously and therefore the possibility to probe the parity lifetime in such a 
setup has also not been discovered.
As we will show, even a small coupling between the two MBS strongly suppresses the current. A finite 
parity relaxation rate partially restores the current but qualitatively changes the transport spectrum, allowing
a finite lifetime to be detected, and in addition its precise value can be measured if the MBS coupling 
or tunnel rates can be controlled.

We solve exactly the problem of a strongly interacting 
quantum dot coupled to a MBS and treat the coupling to the normal metal perturbatively, yielding a set of 
master equations for the state of the combined dot--MBS system.
From these equations we 
calculate the non-linear current as a function of the bias voltage applied to the normal lead and of the gate voltage 
which controls the quantum dot energy. 
In addition, in the supporting information we provide an exact solution of the transport problem 
for the case of a non-interacting quantum dot~\cite{EPAPS}.

The basic properties of MBS in semiconducting systems with induced superconductivity have been 
described in e.g., Refs.~\cite{Oreg10, Alicea10, Lutchyn10, Sau10}.
For suitable parameters, MBS are formed at the ends of the wire, see Fig.~\ref{fig:1}. 
These are zero-energy solutions to the Boguliubov de Genne equations and are 
described by the operators
\begin{eqnarray}\label{eq:MBS}
	\gamma_m	&=& 	\int d z \sum_\sigma \left( g_{\sigma m} (z) \Psi_\sigma(z) + 
				g_{\sigma m}^{*}(z) \Psi_\sigma^\dagger(z) \right),
\end{eqnarray}
where $\Psi_\sigma^\dagger(z)$ creates an electron with spin projection $\sigma = \uparrow,\downarrow$ at 
position $z$ in the wire. 
The exact form of the envelope functions $g_{\sigma m} (z)$, describing the spatial form of the MBS, depends on the details 
of the wire, but $g_{\sigma 1}$ and $g_{\sigma 2}$ have their main weight close to opposite ends and 
decay exponentially inside the wire.
Clearly, the MBS operators satisfy $\gamma_m = \gamma_m^\dagger$ and we assume them 
to be normalized, such that $\gamma_1 \gamma_2 = - \gamma_2 \gamma_1$, $\gamma_m^2 = 1$.
One end of the wire is tunnel coupled to a quantum dot. The coupled dot--MBS system is described by
\begin{eqnarray}\label{eq:H_QD_MBS}
	H_0	&=& 	H_D + \frac{i}{2} \xi \gamma_1 \gamma_2 + \sum_\sigma \left( \lambda_\sigma d_\sigma - 
			\lambda_\sigma^{*} d_\sigma^\dagger \right) \gamma_1,
\end{eqnarray}
where $H_D = \sum_\sigma \epsilon_\sigma n_\sigma + U n_{\uparrow} n_{\downarrow}$ describes the dot, 
$n_\sigma = d_\sigma^\dagger d_\sigma$ is the number operator, and $U$ is the Coulomb charging energy for electrons on the dot.
The low-energy Hamiltonian describing the two MBS was discussed in Ref.~\cite{Kitaev01} and has been used 
in a transport setup in e.g., Refs.~\cite{Tewari08, Lutchyn10, Bolech07, Flensberg10, Law09};
the coupling of the two MBS 
is given by $\xi$,  and $\lambda_\sigma$ is the coupling between electrons on the dot and 
the MBS at the corresponding end of the wire. To be specific, we assume the wire to have a Rashba-type spin-orbit coupling 
along the $x$-direction, and apply a magnetic field $B$ along $z$, leading to a
Zeeman splitting on the dot, $\Delta_B = \epsilon_\uparrow - \epsilon_\downarrow$.
For definiteness, we take the MBS wave function at the dot side of the wire to be
$\propto \Psi_{\downarrow_y}^\dagger(z) + \Psi_{\downarrow_y}(z)$ (see e.g.,~\cite{Oreg10}), 
where $\downarrow_y$ means spin along the negative $y$-axis,
and therefore couples only to the 
$y$-component of the dot spin, equal in amplitude but different in phase for the two dot spin directions, 
$\lambda_\uparrow = \lambda$, $\lambda_\downarrow = -i \lambda$.
The metallic normal lead is described by $H_N = \sum_{k \sigma} \epsilon_{k} c_{k \sigma}^\dagger c_{k \sigma}$ and it couples to the 
dot via $H_T = \sum_{k \sigma} t_{k} d_\sigma c_{k \sigma}^\dagger + h. c.$,
where $c_{k \sigma}^\dagger$ creates an electron in the normal lead with 
energy $\epsilon_{k}$, and where $t_{k}$ is the amplitude for dot$\leftrightarrow$normal lead tunneling.

It is useful to switch to a representation where the two 
Majorana Fermions are combined to form one ordinary Fermion:
$\gamma_1 = f + f^\dagger$, $\gamma_2 = i(f^\dagger - f)$, where $f^\dagger$ creates a Fermion and 
$f^\dagger f = 0,1$ counts the occupation of the corresponding state. The Hamiltonian~(\ref{eq:H_QD_MBS}) now becomes
\begin{eqnarray}\label{eq:H_QD_MBS_fermion}
	H_0	&=& 	H_D + \xi \left( f^\dagger f  - \frac{1}{2} \right) \nonumber \\ 
		&+&	\sum_\sigma \left( \lambda_\sigma d_\sigma f^\dagger
			+ \lambda_\sigma d_\sigma f + h. c. \right), 
\end{eqnarray}
When $\lambda = 0$, the eigenstates 
of~(\ref{eq:H_QD_MBS_fermion}) are given by $|n_d n_f\rangle$, where $n_d = 0, \uparrow, \downarrow, 2$ and $n_f = 0,1$
describe the states of the dot and the MBS system, respectively. Also when $\lambda \neq 0$, 
the Hamiltonian~(\ref{eq:H_QD_MBS_fermion}) is block-diagonal in this basis, with an even block $H_0^{(e)}$, acting on 
$|0 0\rangle, |\sigma 1\rangle, |2 0\rangle$, and an odd block $H_0^{(o)}$, acting on 
$|\sigma 0\rangle, |0 1\rangle, |2 1\rangle$. 
Fermion number is not conserved by~(\ref{eq:H_QD_MBS_fermion}), but, because of the block-structure, 
the \emph{parity} of the \emph{total} Fermion number of the dot--MBS system is conserved.
When $\xi = 0$, $H_0^{(e)}$ and $H_0^{(o)}$ are identical, $\xi \neq 0$ breaks this parity symmetry.

Diagonalizing~(\ref{eq:H_QD_MBS_fermion}) yields odd/even parity eigenstates 
$|o_i /e_i \rangle = \sum_{n_d n_f\in o/e} \alpha^{o_i/e_i}_{n_d n_f} |n_d n_f \rangle$. 
The tunnel Hamiltonian, $H_T$, changes the dot electron number by $\pm 1$ and thus connects 
the even and odd parity sections: 
$\langle e_i | H_T | o_j \rangle = \sum_{k \sigma} K_{k \sigma}^{e_i o_j} c_{k \sigma}^\dagger + h.c.$, where the many-body tunnel
matrix elements are given by
\begin{align}\label{eq:TME}
	K_{k \sigma}^{e_i o_j}	&=& 	t_{k} \sum_{\substack{n_d n_f\in e \\ n_d' n_f'\in o}} \alpha^{e_i *}_{n_d n_f} 
					\alpha^{o_j}_{n_d' n_f'} \langle n_d n_f | d_\sigma | n_d' n_f' \rangle. 
\end{align}
In the presence of strong electron--electron interactions, the tunneling between the normal lead and the dot--MBS system cannot 
be solved exactly (in the limit $U = 0$ we provide an exact solution in the supporting information~\cite{EPAPS}). 
To leading order in $H_T$, the problem can be formulated in terms of master equations~\cite{Bruus04book}, 
which we modify to account for the lack of 
particle conservation. 
The occupations $P_a$ of the dot--MBS eigenstates $|a\rangle = |e_i\rangle , |o_i\rangle$ 
are calculated from 
\begin{eqnarray}\label{eq:ME}
	0 &=&	\sum_{a'} \left( W_{a a'} P_{a'} - W_{a' a} P_{a}\right), \;\; \sum_a P_a = 1, \\
\label{eq:rate_matrix}
	W_{a a'} &=& 	\sum_{\sigma} \left\{ \Gamma_{\sigma}^{a a'} F\left(E_{a a'}\right) + \Gamma_{\sigma}^{a' a} 
			\left[1-F\left(E_{a' a}\right)\right]\right\},
\end{eqnarray}
where $E_{a a'} = E_a - E_{a'}$, with $E_a$ the energy of eigenstate $|a\rangle$, and $F(E) = 1/(e^{(E-\mu_N)/T} + 1)$ is 
the Fermi function of the normal lead with electron temperature $T$ and chemical potential $\mu_N = V$.
(We set $k_B = e = \hbar = 1$.)
The tunnel couplings are $\Gamma_{\sigma}^{a a'} = 2 \pi \rho_N |K_{\sigma}^{a a'}|^2$, with $\rho_N$
being the density of states in the normal lead. 
Note that due to the lack of particle conservation, any pair of states $a, a'$ are connected both by processes 
removing and adding an electron to the normal lead, described by the first and second term in Eq.~(\ref{eq:rate_matrix})
respectively.
We here assumed energy independent density of states and tunnel amplitude, but
the eigenstates $a,a'$ depend on the dot level position and the effective tunnel couplings 
$\Gamma_{\sigma}^{a a'}$ vary between $0$ and $2 \pi \rho_N |t|^2 = \Gamma$.
The physics described by Eq.~(\ref{eq:ME}) is easily understood. The occupation probability of state $a$ is given by the sum of
all tunnel processes starting from any state $a'$ and ending with occupation of state $a$ (first term), minus all processes
depopulating state $a$ (second term). The occupations are normalized to one.
Note that the rates in Eq.~(\ref{eq:rate_matrix}) describe electron tunneling only, rates related to parity relaxation can be added 
to $W_{a a'}$ as described below.
The particle current flowing into the dot from the normal lead is given by 
$I = \sum_{a a'} W_{a a'}^I P_{a'}$,
where the current rate matrix $ W_{a a'}^I$ is similar to~(\ref{eq:rate_matrix}), but with a minus sign on the second term, corresponding 
to electrons tunneling into the normal lead. 

The restriction to lowest order perturbation theory is valid when $T \gg \Gamma$.  
In addition, we neglected off-diagonal elements of the density matrix, valid when all eigenstates are well
separated on a scale set by the tunnel broadening, satisfied when $\lambda \gg \Gamma$. 
Neglecting all states in the superconductor except the MBS is valid when
the induced superconducting gap $\Delta$ is larger than the other energy scales.
Moreover, $U$ can be several meV in small quantum dots formed in semiconducting wires~\cite{Jespersen06, Nilsson09}, 
which is likely $\gtrsim \Delta$, and the same holds for the level-spacing, motivating the single-orbital model used here.

\begin{figure}[t!]
\includegraphics[height=0.9\linewidth]{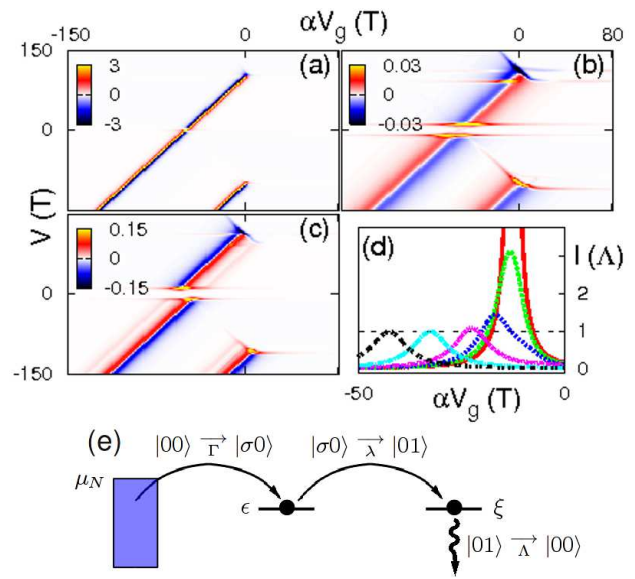}	
	\caption{\label{fig:2}
        (Color online) Stability diagrams: $dI/dV$ normalized by $ 0.02 e^2 / h$, 
	plotted on color scale as function of $V$ and $V_g$, with 
	(a): $\xi = 0$, 
	(b): $\xi = 10T$,
	(c): $\xi = 10T$ and finite parity relaxation, $\Lambda = 0.04 \Gamma$.
	Note the different scales in (b) and (c) showing the dramatic effect of the finite parity lifetime
	and the characteristic shift in gate voltage position between the positive and negative bias peaks in (c).
	(d): $I(V_g)$ at $V = 75 T$ with parameters as in (c), but increasing $\xi / \lambda = 0, 2.5, 5, 10, 20, 30$ 
	(increasing as indicated by the arrow). 
	For $\Lambda \gg \Gamma \lambda^2 / \xi^2$, $\Lambda$ can be read off directly from the height of the shifted 
	current peaks.
	(e): Sketch of resonant transport processes for $\epsilon = \xi$ when $V>0$ in (c).
	Similar resonant processes exist for $\epsilon = -\xi$ when $V < 0$.}
\end{figure}

Next, we present solutions of the master equations for specific choices of parameters, and with the above energy scales 
and limitations in mind, we take $U = \infty$ and $T = \lambda = 10 \Gamma = \Delta_B / 100$, with $T \lesssim 100$~mK.
The voltage-dependence of the energy cost for adding an electron to the dot can be included in the orbital energies,
$(\epsilon_\downarrow + \epsilon_\uparrow)/2 = \epsilon_0 -\alpha V_g + \beta V $, 
where $\alpha$ (gate coupling) and $\beta$ depend on the capacitances associated with the dot--superconductor 
and dot--normal metal tunnel junctions.
We take $\beta = 1/2$ and $\epsilon_0 = 0$ (other choices affect only the slope and absolute position of 
resonances, respectively).
The differential conductance, $dI/dV$, plotted on color scale as function of $V$ and $V_g$ ("stability diagram"), 
is shown in Fig.~\ref{fig:2} for different values of $\lambda$ and $\xi$.

To develop an understanding of the physics underlaying the stability diagrams, we consider first the case without 
parity relaxation, described by Eqs.~(\ref{eq:ME}) and~(\ref{eq:rate_matrix}).
We start with the simplest case of $\xi = 0$, see Fig.~\ref{fig:2}(a). 
A significant current can only flow when one of the dot
spin states is energetically aligned with the MBS, 
allowing electrons to tunnel resonantly between the dot and superconductor, 
resulting in a current peak of width $\sim \lambda$ and
height $I_\text{max} = \Gamma / 2$ and therefore in narrowly spaced positive 
and negative conductance lines. Thus, the sharp nature of the MBS gives rise to negative differential resistance 
in transport through the dot. 
The peak associated with $\sigma = \uparrow$ terminates at large positive $V$, 
when an electron can tunnel from the normal lead into the $\sigma = \downarrow$ state,
blocking transport through the $\sigma = \uparrow$ channel 
(via the Coulomb charging energy)
since it cannot tunnel out 
again into the normal lead (due to energy conservation) or into the MBS (due to the energy mis-match).
Similarly, the $\sigma = \downarrow$ peak vanishes above a negative threshold $V$, when $\sigma = \uparrow$ falls below $\mu_N$.
Thus, properly accounting for Coulomb blockade is essential to understand the qualitative 
conductance features.
In Fig.~\ref{fig:2}(b),  
a finite $\xi$ introduces an energy splitting between the even and odd parity sectors,
suppressing the current for $|V| < \xi$. Even for $|V| > \xi$ the conductance is significantly lower compared to $\xi = 0$ 
[note the different scales in Figs.~\ref{fig:2}(a) and (b)]. To understand why, consider e.g., electrons being 
transported from the normal lead to the superconductor ($V > 0$) by sequential tunneling through the dot, 
which in the basis of the unperturbed 
number states $|n_d n_f\rangle$ involves the processes: 
$|0 0\rangle \; \substack{\longrightarrow \\ \Gamma} \; |\sigma 0\rangle 
\; \substack{\longrightarrow \\ \lambda} \; |0 1\rangle 
\; \substack{\longrightarrow \\ \Gamma} \; |\sigma 1\rangle 
\; \substack{\longrightarrow \\ \lambda} \; |0 0\rangle$.
Because we consider here the case without parity relaxation in the superconductor, any process changing 
the parity of the dot--MBS system must involve tunneling to/from the normal lead, and therefore
two electrons are transferred before the system returns to its initial state. When the even--odd 
degeneracy is split by $\xi \neq 0$, both dot--MBS tunnel processes cannot be resonant at the same voltage,
suppressing the conductance by a factor $\propto \lambda^2 / \xi^2$.
Instead, horizontal conductance lines appear, corresponding 
to inelastic cotunneling~\cite{DeFranceschi01} through the dot,
exciting the parity degree of freedom, i.e., a direct processes
$|0 0\rangle \; \substack{\longrightarrow \\ \Gamma \lambda} \; |0 1\rangle$,
energetically allowed for $|V| > \xi$ and 
involving only virtual occupation of the intermediate state $|\sigma 0\rangle$.
Note that all tunnel process 
$\propto \Gamma \lambda^n$ are included in our master equations, since the dot--MBS coupling is treated exactly.
Interestingly, due to the sharp MBS, inelastic cotunneling gives rise to conductance peaks, rather than steps.
Similarly, horizontal lines at $V = \pm \Delta_B$ correspond to inelastic cotunneling exciting the dot, e.g., 
$|\uparrow 0\rangle \; \substack{\longrightarrow \\ \lambda \Gamma } \; |\downarrow 0\rangle$.

Next, we introduce a finite relaxation of the Fermion parity in the superconductor, 
caused by quasiparticle poisoning~\cite{Mannik04, Aumentado04}, 
and show that this has a striking effect on the stability diagrams.
Rather than starting from a microscopic model of the quasiparticle generation, 
we focus on the generic effects and assume that transitions 
between dot--MBS eigenstates $|a' \rangle \rightarrow | a \rangle$ are induced with rate 
$\Lambda_{a a'}$ if $E_{a'} > E_{a}$ and rate $\Lambda_{a a'} \text{exp}[(E_{a'} - E_{a})/T] $ otherwise, 
where $\Lambda_{a a'}$ is obtained by assuming a rate $\Lambda$ for transitions
$|n_d n_f \rangle \rightarrow |n_d \bar{n}_f \rangle$ ($\bar{1}=0, \bar{0}=1$), and then transforming 
to the dot--MBS eigenbasis, cf., Eq.~(\ref{eq:TME}). These rates are then added to the rate matrix
in Eq.~(\ref{eq:rate_matrix}).
As is seen by comparing Fig.~\ref{fig:2}(c) and (b), which differ only by a small such relaxation rate,
this has a dramatic effect on the conductance (note the different scales).
To understand the increased conductance, consider transport at $V > 0$,
which for $\Lambda \neq 0$ is dominated by the processes:
$|0 0\rangle \; \substack{\longrightarrow \\ \Gamma} \; |\sigma 0\rangle 
\; \substack{\longrightarrow \\ \lambda} \; |0 1\rangle 
\; \substack{\longrightarrow \\ \Lambda} \; |0 0\rangle$, see sketch in Fig.~\ref{fig:2}(e).
Hence, parity relaxation allows the total dot--MBS system to return to its initial state 
after transferring only a single electron, and this electron tunnels through the system fully resonantly 
when $\epsilon_\sigma = \xi$.
Similarly, at $V < 0$ transport is dominated by:
$|0 0\rangle \; \substack{\longrightarrow \\ \lambda} \; |\sigma 1\rangle 
\; \substack{\longrightarrow \\ \Lambda} \; |\sigma 0\rangle 
\; \substack{\longrightarrow \\ \Gamma} \; |0 0\rangle$, 
which is fully resonant at $\epsilon_\sigma = -\xi$.
Thus, parity relaxation leads to resonant transport even for finite $\xi$, but the resonance condition is different for 
positive and negative bias and therefore the peaks are shifted by $2 \xi$.

The shifted peak behavior is observed when 
the parity relaxation rate exceeds the suppressed transport rates discussed in 
connection with Fig.~\ref{fig:2}(b), i.e., when
$\Lambda \gtrsim \Gamma \lambda^2 / \xi^2 $.
If, in addition, $\Gamma \gg \Lambda$, parity relaxation is the slowest process in Fig.~\ref{fig:2}(e) and 
the peak current is equal to $\Lambda$~\cite{maxcurrent}.
Experimentally, $\xi$ could be changed by moving the MBS corresponding to $\gamma_2$ 
with "keyboard" gates as suggested in Ref.~\cite{Alicea10b}. $I(V_g)$ curves for increasing 
$\xi$ are shown in Fig.~\ref{fig:2}(d), allowing the parity relaxation to be directly read off 
from the height of the  current peaks at large $\xi$. 
Alternatively, one could change instead $\Gamma$ and/or $\lambda$ via additional gates.
Note that $\Lambda$ can be directly measured without knowing the precise values of e.g., $\Gamma$ and $\lambda$.
We emphasize that even at finite $\xi$, the two-Majorana state described by $\gamma_{1,2}$ or $f^\dagger, f$ can 
only relax by also changing the parity of the superconductor. Therefore, even though our proposed measurement has 
to be done at finite $\xi$, the relaxation rate measured is the parity relaxation rate, which is arguably independent 
of $\xi$.

In conclusion, we have suggested a relatively simple setup in which to measure the lifetime of the parity of
a Majorana system, being of crucial importance in topological quantum computing schemes~\cite{Nayak08rev}. 
Our suggestion does not involve (experimentally problematic) interferometry or braiding operations.
Thus, one can test different realizations of topological superconductors and different material choices, selecting the most 
suitable ones, \emph{before} going through the difficult process of actually constructing a topological quantum computer.
The setup we propose is a normal metal--quantum dot--topological superconductor junction. We
have calculated the non-linear conductance of this system as a function of the applied gate and bias voltages within 
a master equation approach and shown that the lifetime of the parity of the Majorana system 
can be directly read off from the height of the current peaks at finite Majorana coupling.

We thank Harvard University, where part of this work was done, for hospitality. 
This work was funded in part by The Danish Council for Independent Research $|$ Natural Sciences and by Microsoft Corporation Project Q.

\bibliographystyle{apsrev}

\begin{thebibliography}{10}
\bibitem{maxcurrent}%
  \BibitemOpen
  \bibinfo {note} {For $\Gamma \gg \Lambda \gg \Gamma \lambda^2 / \xi^2$, the
  peak current expanded in $\Lambda / \Gamma$ and $\lambda / \xi$ is
  $I_\text{max} \approx \Lambda (1 + \frac{1}{4}
  \frac{\Gamma}{\Lambda}\frac{\lambda^2}{\xi^2} - 3\frac{\Lambda}{\Gamma})$}%
  \bibAnnoteFile{NoStop}{maxcurrent}%
\end{thebibliography}
\section*{Supplementary information}
\begin{figure}[h!]
  \includegraphics[height=0.41\linewidth]{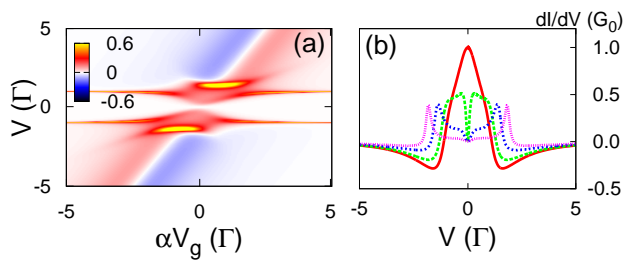}	
	\caption{\label{fig:3}
        (Color online) (a) $dI/dV$ normalized by $2 e^2 / h$, plotted on color scale as function of $V$ and $V_g$, 
	with $\xi = \Gamma = 2 \lambda $, $T = 0$. We take the voltage coupling $\beta = 1/2$ as in 
	Fig.~2 in the main paper, but choose the  
	gate voltage scale such	that $\epsilon_\uparrow = 0$ when $V_g = V = 0$.
        (b) $dI/dV$ as function of $V$ at $V_g = 0$. The parameters are the same as in (a), except for $\xi$ which is varied: 
	$\xi = 0$ (red solid line), $\xi = \Gamma/2$ (green dashed line), $\xi = \Gamma$ (blue dotted line), 
	$\xi = 3 \Gamma/2$ (purple fine dotted line). 
	}
\end{figure}
We here consider the Zeeman splitting to be the largest energy scale, 
$\Delta_B \gg |V|,T,\Gamma,\lambda$, in which case 
transport through the ground state of the dot can be described by a non-interacting model and solved 
exactly using a Green's function approach.
The current can be written as
\begin{eqnarray}\label{eq:GFcurrent4}
	I	&=&		-\Gamma \int \frac{d\omega}{2 \pi} \; \left[ 2 \text{Im} G^R_{d d} + 
				\Gamma \left( |G^R_{d d}|^2 - | F^R_{d d} |^2 \right) \right] \nonumber \\ 
		&\times&	\left[ F(\omega) - F_0(\omega) \right],
\end{eqnarray}
where $F_0(\omega)$ is the Fermi function at $V=0$, $G^R_{d d}$ is the Fourier transform of the retarded dot Green's function
for $\sigma = \uparrow$ electrons, and 
$F^R_{d d} (t - t') = -i \theta(t - t') \langle \{ d_\uparrow(t), d_\uparrow(t') \} \rangle$ 
is the corresponding anomalous Green's function. (We here only consider $\Lambda = 0$.)
Next, we take the time derivative of $G^R_{d d}(t - t')$ [and of all Green's functions appearing when doing so, including 
$F^R_{d d}(t - t')$], Fourier transform, and solve the resulting equations of motion, which gives
\begin{align}\label{eq:GRd}
	G^R_{d d}	=	\frac{1}{ \omega - \epsilon_\uparrow + \frac{i}{2}\Gamma - 2 |\lambda|^2 M(\omega) 
				\left[ 1 + 2 |\lambda|^2 \tilde{M}(\omega)\right] },
\end{align}
\begin{multline}
\label{eq:GANd}
	F^R_{d d}	=	\\ \frac{-2 (\lambda^{*})^2 M(\omega)}{\left( \omega + \epsilon_\uparrow + \frac{i}{2}\Gamma \right) 
								\left( \omega - \epsilon_\uparrow + \frac{i}{2}\Gamma \right) - 4 |\lambda|^2 
								\left( \omega + \frac{i}{2}\Gamma \right) M(\omega)},
\end{multline}
where $M(\omega) = 1/ (\omega - \xi^2/\omega)$, and 
$\tilde{M}(\omega) = M(\omega) / [\omega + \epsilon_\uparrow + \frac{i}{2}\Gamma - 2 |\lambda|^2 M(\omega)]$.

When $\lambda, T \gg \Gamma$, the results agree fully with 
the master equation results for $B = \infty$ and we now focus instead on the limit $T = 0, \lambda \lesssim \Gamma$.
Figure~\ref{fig:3}(a) shows a stability diagram similar to those in Fig.~2 in the main paper, but for a smaller voltage window 
close to the $\epsilon_\uparrow = 0$ crossing. 

Figure~\ref{fig:3}(b) shows $dI/dV$ as function of $V$ at $V_g = 0$, for varying values of $\xi$.  
The zero-bias conductance, $G(0)$, can be found analytically from~(\ref{eq:GFcurrent4}).
When $\xi=0$,  we always have $G(0) = 2 e^2 / h$, independent of $V_g$, although the width of the 
peak decreases when the dot level is brought out of resonance. Conversely, for any $\xi \neq 0$, we find $G(0) = 0$, independent 
of $V_g$. This is similar to the linear conductance of a normal metal--MBS tunnel junction,
see Ref.~[16] of the main paper.
At finite $\xi$, peaks instead appear at finite voltage $V \approx \pm \xi$, as discussed in connection with the 
master equation results in the main paper, and close to $V_g = 0$ an interesting double peak structure is found. 
\end{document}